\documentclass{webofc}
\usepackage[varg]{txfonts} 

\begin{document}

\title{Coupled baryon, electric charge and strangeness fluctuations in heavy-ion collisions}

\author{\firstname{Grégoire} \lastname{Pihan}\inst{1}\fnsep\thanks{\email{pihan@subatech.in2p3.fr}} \and
        \firstname{Marcus} \lastname{Bluhm}\inst{1}
        \and
        \firstname{Marlene} \lastname{Nahrgang}\inst{1}
}

\institute{SUBATECH UMR 6457 (IMT Atlantique, Université de Nantes, IN2P3/CNRS), 4 rue Alfred Kastler, 44307 Nantes, France}

\abstract{
    We present for the first time quantitative results for the coupled dynamics of second order fluctuations in the three conserved charges of QCD based on stochastic diffusion equations for a Bjorken-type expanding hadronic medium. 
  }
\maketitle
\section{Introduction}
\label{sec-1}
Fluctuation observables in heavy-ion collisions test the constituents and the transport properties of strongly interacting matter and may signal phase transitions or the chemical freeze-out. The fluctuations in the three conserved charges of QCD, net-baryon number $B$, net-electric charge $Q$ and net-strangeness $S$, change through diffusion~\cite{Asakawa:2019kek}. Their evolution is coupled as the matter constituents typically carry more than just one charge. In this work we study the coupled fluctuation dynamics in the hadronic phase employing stochastic diffusion equations. These are expressed in terms of the Milne coordinates proper-time $\tau$ and spatial rapidity $y$ to model a rapidly Bjorken-type expanding medium. By using realistic values for the diffusion coefficients we investigate the competition between coupled diffusion and rapid expansion and find that the fluctuation observables are necessarily driven out of equilibrium. This may have important phenomenological consequences for the interpretation of experimental data.

\section{Coupled conserved charge diffusion in the hadronic phase}
\label{sec-2}
The dynamics of the net-charges $i=B,Q,S$ is influenced by the mixing of different diffusive currents entering the stochastic diffusion equations 
\begin{equation}
    \partial_\tau n_i = \sum_{j=B,Q,S} \partial_y \left(
    \frac{\kappa_{ij}}{\tau} \partial_y \left(\frac{\mu_j}{T}\right)
    \right)
    + \partial_y \xi_i \,.
\label{equ:mixeddiffusion}
\end{equation}
Here, $\kappa_{ij}$ represents the charge diffusion coefficient matrix of the hadronic medium. 

The fluctuation-dissipation balance requires that the noise satisfies besides the standard white-noise self-correlation also cross-correlations between the different charges introduced by the non-vanishing off-diagonal elements of $\kappa_{ij}$. These are given, for $Y=(\tau,y)$, by 
\begin{equation}
    \langle \xi_i (Y)\,\xi_j(Y') \rangle = 2 \frac{\kappa_{ij}}{\tau} \delta^{(2)}(Y-Y') \,.
\label{equ:crosscorrelations}
\end{equation}
We note that charge coupling may not only be established by means of the current mixing and the noise cross-correlations but also via the underlying equation of state for which $\mu_j$ may depend on $\mu_{i\ne j}$. In~\cite{Fotakis:2019nbq}, the deterministic diffusion of the net-charge densities $n_i$ was studied. 
It was found that the charge couplings affect the density profiles non-trivially. The current work presents the first approach to study the coupled dynamics of conserved charge fluctuations quantitatively by means of the set of stochastic diffusion equations~(\ref{equ:mixeddiffusion}) and~(\ref{equ:crosscorrelations}). 
\section{Numerical results}
\label{sec-3}
\begin{figure}[b!]
    \centering
    \includegraphics[width=0.44\textwidth,clip]{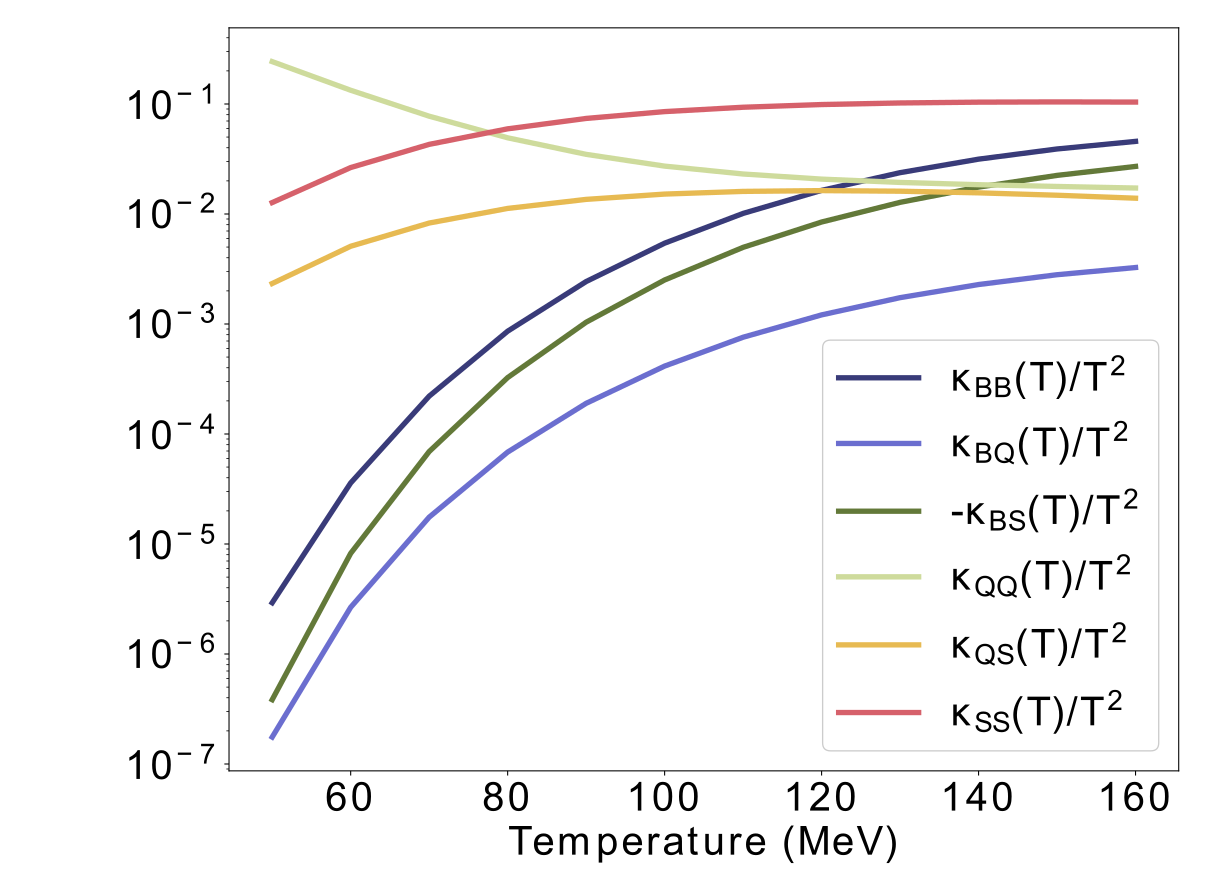}
    \hspace{0.03\textwidth}
    \includegraphics[width=0.45\textwidth,clip]{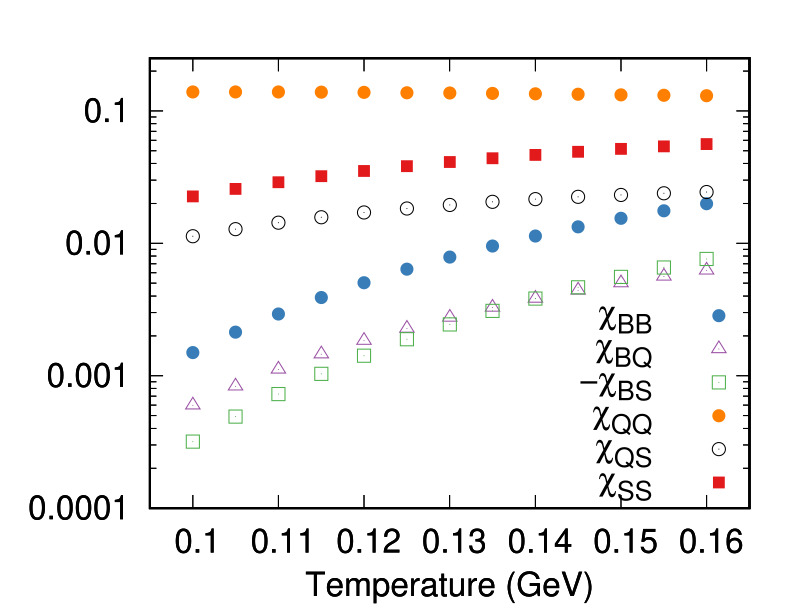}
    \caption{Left: Temperature dependence of the diffusion coefficients $\kappa_{ij}$ at $\mu_B=0$~MeV for a hadron resonance gas composed of the 19 lightest hadron species (taken from~\cite{Greif:2017byw}). Right: Corresponding equilibrium diagonal and off-diagonal susceptibilities per unit of rapidity as function of $T$ at $\mu_B=0$~MeV.}
    \label{fig:kij}
\end{figure}
The diffusion coefficient matrix $\kappa_{ij}$ as function of temperature $T$ and chemical potential $\mu_j$ has been calculated within kinetic theory in~\cite{Greif:2017byw} for a hadron resonance gas composed of the 19 lightest hadron species. Consequently, we employ the same underlying model in order to determine the spatio-temporal profile of the different $\mu_j$ in Eqs.~(\ref{equ:mixeddiffusion}) from the charge density profiles assuming a homogeneous $T$. 
For the latter, the temporal evolution is modeled Hubble-like viz $T(\tau)=T(\tau_0)\tau_0/\tau$ with $\tau_0=1$~fm$/$c and $T(\tau_0)=500$~MeV. Initially, we set $n_i(\tau,y)=0$~fm$^{-2}$ 
which implies $\mu_j=0$~MeV for the chemical potential profiles. In Fig.~\ref{fig:kij} (left panel), we show the $T$-dependence of $\kappa_{ij}$ at $\mu_B=0$~MeV. In addition, the initial local variances and co-variances are also set to vanish, clearly representing a far from equilibrium situation. The imposed balance between fluctuation and dissipation allows the system to approach equilibrium, but based on the strength of the relevant $\kappa_{ij}$ a rapid expansion may compete with this process. In Fig.~\ref{fig:kij} (right panel), we show the expected $T$-dependence of the equilibrium fluctuations of the hadronic medium.

\subsection{Equilibration process}
\label{sec-3a}
\begin{figure}[t!]
    \centering
    \includegraphics[width=0.45\textwidth,clip]{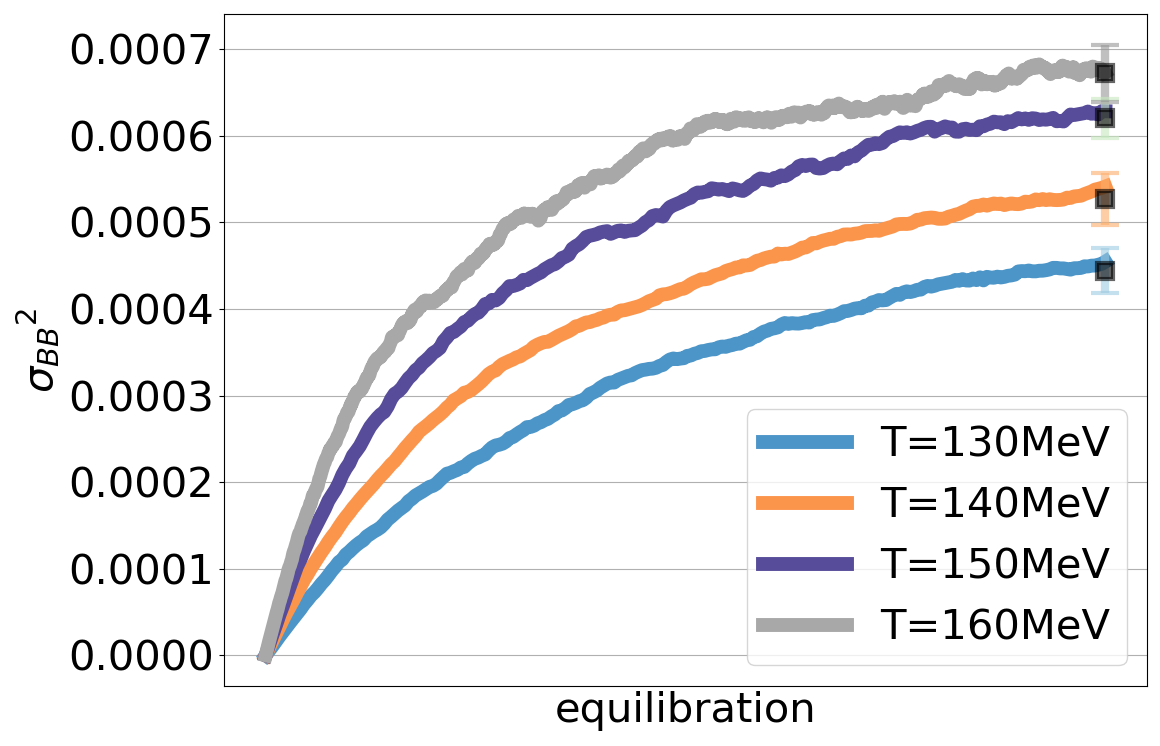}
    \hspace{0.03\textwidth}
    \includegraphics[width=0.45\textwidth,clip]{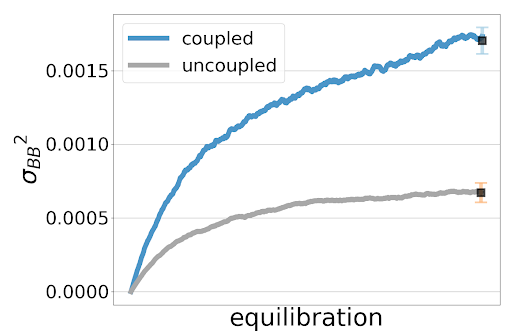}
    \caption{Time-evolution of $\sigma_{BB}^2$ during the equilibration process for $B$ not coupled to $Q,S$ at different $T$ (left) and in comparison to the fully coupled system (right) at $T=160$~MeV. The squares show our values for the equilibrium fluctuations with errors estimated through the late-time slope of the curves.
    }
    \label{fig:equilibration}
\end{figure}

The success of the fluid dynamic modeling of heavy-ion collisions suggests that the created bulk matter is locally near equilibrium. For the given realistic values of $\kappa_{ij}$~\cite{Greif:2017byw} it is, thus, important to understand how quickly equilibrium can be achieved starting from our far-from-equilibrium initial conditions. To this end, Eqs.~(\ref{equ:mixeddiffusion}) and~(\ref{equ:crosscorrelations}) are solved at fixed $T$ for a static system. In this way we can create an equilibrated initial state for a subsequent dynamic expansion as well as provide estimates for the equilibrium fluctuations realized at a given $T$. In Fig.~\ref{fig:equilibration}, we show the equilibration process for the local variance of the net-baryon density, $\sigma_{BB}^2$, in the limit $B$ not coupled to $Q,S$ (left panel) for different $T$ and for the fully coupled system (right panel) in comparison to the uncoupled situation for $T=160$~MeV. One observes that for the given values of $\kappa_{ij}$ equilibrium is only slowly reached (the waited equilibration time is of $\mathcal{O}(20)$~fm$/$c), in particular for smaller $T$ in line with the $T$-dependence of $\kappa_{BB}$. Moreover, our equilibrium estimates for $\sigma_{BB}^2$ behave qualitatively similar to $\chi_{BB}$, see Fig.~\ref{fig:kij}. The coupling between the different charges leads to an even slower approach of local equilibrium, where the size of the equilibrium fluctuations is increased compared to the uncoupled situation.

\subsection{Dynamic expansion and cooling}
\label{sec-3b}
\begin{figure}[t!]
    \centering
    \includegraphics[width=0.45\textwidth,clip]{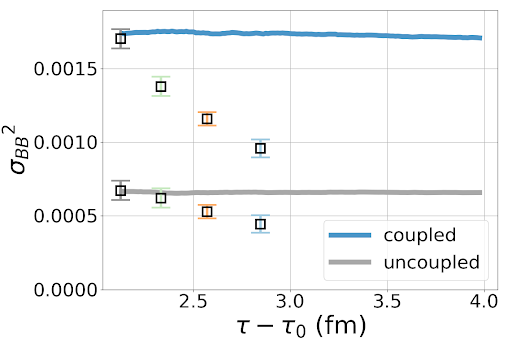}
    \hspace{0.03\textwidth}
    \includegraphics[width=0.45\textwidth,clip]{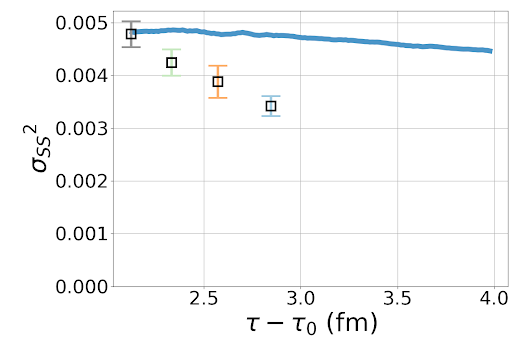}
    \caption{Nonequilibrium time-evolution of $\sigma_{BB}^2$ (left) and $\sigma_{SS}^2$ (right) in the hadronic phase between $T=160$~MeV and $T=100$~MeV for the fully coupled conserved charge dynamics in comparison with our equilibrium estimates (squares) for both fluctuation observables. For $\sigma_{BB}^2$ we contrast, in addition, the corresponding results obtained in the limit $B$ not coupled to $Q,S$.
}
\label{fig:expansion}
\end{figure}

Starting from the created equilibrium initial conditions at $T=160$~MeV, see Sec.~\ref{sec-3a}, we can study the dynamics of the fluctuation observables as $T(\tau)$ cools down. In Fig.~\ref{fig:expansion}, we show the time-evolution of the local variances $\sigma_{BB}^2$ and $\sigma_{SS}^2$ for the net-baryon (left panel) and net-strangeness (right panel) density from $T=160$~MeV to $T=100$~MeV. In comparison to our estimates for the equilibrium fluctuations at different $T$ (squares) one observes that for the $\kappa_{ij}$ given in~\cite{Greif:2017byw} the conserved charge fluctuations lag behind and are quickly driven out of local equilibrium in the rapidly expanding system. Deviations in  $\sigma_{BB}^2$ from our equilibrium expectations are found to be stronger in the coupled system as the simultaneous nonequilibrium evolution in $Q$ and $S$ affects the dynamics of the fluctuations in $B$. In addition, we find that $\sigma_{SS}^2$ follows the equilibrium situation better, mostly because $\kappa_{SS}$ is the largest diffusion coefficient in the considered temperature range, cf.~Fig.~\ref{fig:kij}.

\section{Discussion, conclusions and outlook}
\label{sec-4}
We have seen that the coupling between the conserved charges in QCD impacts the diffusion dynamics of their fluctuations $\sigma_{BB}^2$ and $\sigma_{SS}^2$ in the hadronic phase significantly. Charge coupling does not only increase the size of fluctuations with possible consequences for the search of the QCD critical point, it also affects to which extent local equilibrium can be achieved and maintained. Starting from an initial condition near local equilibrium at $T=160$~MeV we have shown, using realistic diffusion coefficients $\kappa_{ij}$~\cite{Greif:2017byw}, that it is impossible to maintain equilibrium locally in a Bjorken-type expanding medium. While diffusion still has some impact, the rapid expansion wins and the system is necessarily driven out of equilibrium in the hadronic phase. Charge coupling accelerates this nonequilibrium evolution.

The observed behavior impacts our interpretation of fluctuation measurements: We find that, as a consequence of the diffusive dynamics in the hadronic phase, final fluctuations in the net-baryon number only portray a snapshot of the initial (equilibrium) situation. Moreover, while final net-strangeness fluctuations might still reflect chemical freeze-out conditions these are necessarily over-estimated by the comparison with equilibrium models. This shows that previous determinations as in~\cite{Alba:2014eba,Bluhm:2018aei} based on fluctuation data have to be reassessed. Our study also indicates that the critical fluctuation signals from the QCD phase transition~\cite{Kitazawa:2020kvc} may survive the rapid expansion of the hadronic medium.

In this work, we have not discussed fluctuations in the net-electric charge, which require the largest equilibration time, nor cross-correlations between the different charges. We leave their presentation and a detailed discussion to the future~\cite{Pihan:future}.

\section*{Acknowledgment}
We thank the authors of~\cite{Greif:2017byw} for stimulating discussions and providing us with the data of $\kappa_{ij}$. This work received support from the program "Etoiles montantes en Pays de la Loire 2017".

\end{document}